\documentstyle[12pt]{article}
\begin{document}

\baselineskip 24pt

\newcommand{\sheptitle}
{FERMION MASSES IN SUSY GUTS\footnote{Invited talk
at the International Workshop on Elementary Particle Physics:
Present and Future, Valencia 1995.}}

\newcommand{\shepauthor}
{B. C. Allanach \footnote{Address after 1 October 1995:
Rutherford Appleton Laboratory, Chilton, Didcot, OX11 0QX.}
and S. F. King \footnote{Talk presented by S. F. King.}}

\newcommand{\shepaddress}
{Physics Department, University of Southampton
\\Southampton, SO17 1BJ}

\newcommand{\shepabstract}
{We discuss the fermion
mass problem in SUSY GUTs, including such ideas as
texture zeroes, and Georgi-Jarlskog textures.
We focus on a specific supersymmetric
model based on the gauge group $SU(4)\otimes SU(2)_L \otimes SU(2)_R$.
In this model the gauge group is broken to that of the standard
model at $10^{16}$ GeV, and supersymmetry is broken at low-energy.
The model may be regarded as a ``surrogate SUSY GUT'', and has several
advantages over the $SU(5)$ and $SO(10)$ SUSY GUTs,
such as the absence of the
doublet-triplet splitting problem, and a simpler Higgs sector
which does not involve any large Higgs representations.
The model predicts full quadruple Yukawa unification
(top-bottom-tau-tau neutrino), leading to the
prediction of large top mass and
$\tan \beta$. An operator analysis
leads to schemes where the strange and down mass are predicted,
and $|V_{ub}|>0.004$.}

\begin{titlepage}
\begin{flushright}
SHEP 95-26\\
hep-ph/9508236
\end{flushright}
\vspace{.4in}
\begin{center}
{\large{\bf \sheptitle}}
\bigskip \\ \shepauthor \\ \mbox{}
\\ {\it \shepaddress} \\ \vspace{.5in}
{\bf Abstract} \bigskip \end{center} \setcounter{page}{0}
\shepabstract
\end{titlepage}

\section{Introduction}

The Standard Model involves three gauge couplings $g_i$,
($i=1,2,3$), a Higgs mass parameter $\mu$ and quartic coupling
$\lambda$, and the fermion Yukawa couplings.
In the Standard Model Lagrangian one must specify the
three Yukawa matrices $\lambda^E_{ij}$, $\lambda^D_{ij}$,
$\lambda^U_{ij}$,
corresponding to up to 54 real parameters, which after
diagonalisation lead to 9 physical masses (6 quark masses and 3
charged lepton masses) and 4 physical
quark mixing parameters. Thus the fermion sector of the Standard
Model involves either 54 or 13 unknown free parameters,
depending on how you choose to count them. Either way, from
a fundamental point of view the situation is unacceptable
and the fermion mass problem, as it has been called,
is one motivation for going beyond the Standard Model.
The big question of course is what lies beyond it?

We have not yet experimentally studied the mechanism of electroweak
symmetry breaking, so one might argue that it is premature to study
the fermion mass problem. Unless we can answer this, we have no hope
of understanding anything about fermion masses since we do not have a
starting point from which to analyse the problem.  However LEP has
taught us that whatever breaks electroweak symmetry must do so in a
way which very closely resembles the standard model. This observation
by itself is enough to disfavour many dynamical models involving large
numbers of new fermions.  By contrast the minimal supersymmetric
standard model (MSSM) mimics the standard model very closely.
Furthermore, by accurately measuring the strong coupling constant, LEP
has shown that the gauge couplings of the MSSM merge very accurately
at a scale just above $10^{16}$ GeV, thus providing a hint for
possible unification at this scale. On the theoretical side,
supersymmetry (SUSY) and grand unified theories (GUTs) fit together
very nicely in several ways, providing a solution to the technical
hierarchy problem for example. When SUSY GUTs are extended to
supergravity (SUGRA) the beautiful picture of universal soft SUSY
breaking parameters and radiative electroweak symmetry breaking via a
large top quark yukawa coupling emerges. Finally, there is an on-going
effort to embed all of this structure in superstring models, thereby
allowing a complete unification, including gravity.

Given the promising scenario mentioned above, it is hardly surprising
that many authors have turned to the SUSY GUT framework as a
springboard from which to attack the problem of fermion masses
\cite{Rabyi}.  Indeed in recent years there has been a flood of papers
on fermion masses in SUSY GUTs.  Although the approaches differ in
detail, there are some common successful themes which have been known
for some time. For example the idea of bottom-tau Yukawa unification
in SUSY GUTs \cite{btauold} works well with current data
\cite{btaunew}.  A more ambitious extension of this idea is the
Georgi-Jarlskog (GJ) ansatz (see later)
which provides a successful description of
all down-type quark and charged lepton masses
\cite{GJ}, and which also works well with current data
\cite{DHR}.
In order to understand the origin of the texture zeroes, one must
consider the details of the model at the scale $M_X \sim 10^{16}$
GeV. The alternative is to simply make a list of
assumptions about the nature of the Yukawa matrices at $M_X$
\cite{frogetal}.  For example Ramond, Roberts and Ross (RRR)
\cite{RRR} assumed symmetric Yukawa matrices at $M_X$, together with
the GJ ansatz for the lepton sector.  It is difficult to proceed
beyond this without specifying a particular model. Indeed, this model
dependence may be a good thing since it may mean that the fermion mass
spectrum at low energies is sensitive to the theory at $M_X$, so it
can be used as a window into the high-energy theory.  Therefore in
what follows we shall restrict ourselves to a very specific gauge
group at $M_X$.

Twenty-one years ago Pati and Salam proposed a model in which the
standard model was embedded in the gauge group SU(4)$\otimes$SU(2)$_L
\otimes $SU(2)$_R$ \cite{pati}.  More recently a superpersymmetric
(SUSY) version of this model was proposed in which the gauge group is
broken at $M_{X}\sim 10^{16}$ GeV \cite{leo1}.  The model \cite{leo1}
does not involve adjoint representations and later some attempt was
made to derive it from four-dimensional strings, although there are
some difficulties with the current formulation \cite{leo2}.
The absence of adjoint representations is not an
essential prerequisite for the model to descend from the superstring,
but it leads to some technical simplifications.  Also in the present
model, the colour triplets which are in separate representations
{}from the Higgs doublets, become heavy in a very simple way so the
Higgs doublet-triplet splitting problem does not arise. These two
features (absence of adjoint representations and absence of the
doublet-triplet splitting problem,) are shared by flipped
SU(5)$\otimes $U(1)
\cite{flipped}, which also has a superstring formulation.  Although
the present model and flipped SU(5)$\otimes $U(1) are similar in many
ways, there are some important differences.  Whereas the Yukawa
matrices of flipped SU(5)$\otimes$U(1) are completely unrelated at the
level of the effective field theory at $M_{X}$ (although they may have
relations coming from the string model) in the present model there is
a constraint that the top, bottom and tau Yukawa couplings must all
unify at that scale.  In addition there will be Clebsch relations
between the other elements of the Yukawa matrices, assuming they are
described by non-renormalisable operators, which would not be present
in flipped SU(5)$\otimes $U(1). In these respects the model resembles
the SO(10) model recently analysed by Anderson et al \cite{Larry}.
However it differs from the SO(10) model in that the present model
does not have an SU(5) subgroup which is central to the analysis of
the SO(10) model. In addition the operator structure of the present
model is totally different.  Thus the model under consideration is in
some sense similar to flipped SU(5)$\otimes$U(1), but has third family
Yukawa unification and precise Clebsch relationships as in SO(10). We
find this combination of features quite remarkable, and it seems to us
that this provides a rather strong motivation to study the problem of
fermion masses in this model \cite{422}.

\section{The 422 Model}
Here we briefly summarise the parts of the model
which are relevant for our analysis.
The gauge group is,
\begin{equation}
\mbox{SU(4)}\otimes \mbox{SU(2)}_L \otimes \mbox{SU(2)}_R. \label{422}
\end{equation} The left-handed quarks and leptons are accommodated in
the following representations,
\begin{equation} {F^i}^{\alpha a}=(4,2,1)=
\left(\begin{array}{cccc} u^R & u^B & u^G & \nu \\ d^R & d^B & d^G &
e^-
\end{array} \right)^i
\end{equation}
\begin{equation} {\bar{F}}_{x \alpha}^i=(\bar{4},1,\bar{2})=
\left(\begin{array}{cccc}
\bar{d}^R & \bar{d}^B & \bar{d}^G & e^+ \\
\bar{u}^R & \bar{u}^B & \bar{u}^G & \bar{\nu}
\end{array} \right)^i
\end{equation} where $\alpha=1\ldots 4$ is an SU(4) index, $a,x=1,2$
are SU(2)$_{L,R}$ indices, and $i=1\ldots 3$ is a family index.  The
Higgs fields are contained in the following representations,
\begin{equation} h_{a}^x=(1,\bar{2},2)=
\left(\begin{array}{cc} {h_2}^+ & {h_1}^0 \\ {h_2}^0 & {h_1}^- \\
\end{array} \right) \label{h}
\end{equation} (where $h_1$ and $h_2$ are the low energy Higgs
superfields associated with the MSSM.) The two heavy Higgs
representations are
\begin{equation} {H}^{\alpha b}=(4,1,2)=
\left(\begin{array}{cccc} u_H^R & u_H^B & u_H^G & \nu_H \\ d_H^R &
d_H^B & d_H^G & e_H^-
\end{array} \right) \label{H}
\end{equation} and
\begin{equation} {\bar{H}}_{\alpha x}=(\bar{4},1,\bar{2})=
\left(\begin{array}{cccc}
\bar{d}_H^R & \bar{d}_H^B & \bar{d}_H^G & e_H^+ \\
\bar{u}_H^R & \bar{u}_H^B & \bar{u}_H^G & \bar{\nu}_H
\end{array} \right). \label{barH}
\end{equation}

The Higgs fields are assumed to develop VEVs,
\begin{equation} <H>=<\nu_H>\sim M_{X}, \ \
<\bar{H}>=<\bar{\nu}_H>\sim M_{X}
\label{HVEV}
\end{equation} leading to the symmetry breaking at $M_{X}$
\begin{equation}
\mbox{SU(4)}\otimes \mbox{SU(2)}_L \otimes \mbox{SU(2)}_R
\longrightarrow
\mbox{SU(3)}_C \otimes \mbox{SU(2)}_L \otimes \mbox{U(1)}_Y
\label{422to321}
\end{equation} in the usual notation.  Under the symmetry breaking in
Eq.\ref{422to321}, the Higgs field $h$ in Eq.\ref{h} splits into two
Higgs doublets $h_1$, $h_2$ whose neutral components subsequently
develop weak scale VEVs,
\begin{equation} <h_1^0>=v_1, \ \ <h_2^0>=v_2 \label{vevs}
\end{equation} with $\tan \beta \equiv v_2/v_1$.

Below $M_{X}$ the part of the
superpotential involving quark and charged lepton fields is just
\begin{equation} W
=\lambda^{ij}_UQ_i\bar{U}_jh_2+\lambda^{ij}_DQ_i\bar{D}_jh_1
+\lambda^{ij}_EL_i\bar{E}_jh_1+ \ldots \label{NMSSM}
\end{equation} with the boundary conditions at $M_{X}$,
\begin{equation}
\lambda^{ij}_U=\lambda^{ij}_D=\lambda^{ij}_E.
\label{boundary}
\end{equation}
The same Yukawa relations also occur in minimal $SO(10)$.

\section{The Basic Strategy}

Such Yukawa relations as in Eq.\ref{boundary}
taken at face value
are a phenomenological disaster. For example
consider the minimal $SU(5)$ prediction
$\lambda^{ij}_D=\lambda^{ij}_E$. After diagonalisation
this leads to
$\lambda_e = \lambda_d$,
$\lambda_{\mu} = \lambda_s$,
$\lambda_{\tau} = \lambda_b$,
(at the scale $M_X$)
and hence
\begin{equation}
\frac{\lambda_s}{\lambda_d} = \frac{\lambda_{\mu}}{\lambda_e}
\end{equation}
which is RG invariant and fails badly at low-energy.
On the other hand the third family relation leads to
the low-energy prediction (assuming the SUSY RG equations)
$\lambda_b / \lambda_{\tau} \approx 2.4$ which works well.

A possible fix is provided by the GJ texture,
\begin{eqnarray}
\lambda^E &=& \left(\begin{array}{ccc}
0 & \lambda_{12} & 0 \\
\lambda_{21} & 3 \lambda_{22} & 0 \\
0 & 0 & \lambda_{33} \\ \end{array}\right) \\
\lambda^D &=& \left(\begin{array}{ccc}
0 & \lambda_{12} & 0 \\
\lambda_{21} & \lambda_{22} & 0 \\
0 & 0 & \lambda_{33} \\ \end{array}\right),
\end{eqnarray}
With predictions (at $M_{X}$)
\begin{eqnarray}&&\begin{array}{cc}
\lambda_s = \lambda_\mu/3, & \lambda_d = 3\lambda_e, \\ \end{array}
\end{eqnarray}
which is viable.

As
it turns out, the idea of full top-bottom-tau
Yukawa unification works rather well for the
third family \cite{Yuk}, leading to the prediction of a large top
quark mass $m_t>165$ GeV, and $\tan \beta \sim m_t/m_b$ where $m_b$ is
the bottom quark mass.  However Yukawa unification for the first two
families is not successful, since it would lead to unacceptable mass
relations amongst the lighter fermions, and zero mixing angles at
$M_{X}$. In order to cure these problems, we require
something akin to the GJ texture, in which the Yukawa relations
are altered by group theoretical Clebsch coefficients, leading
to enhanced predictivity.

One interesting proposal has recently been put forward to account for
the fermion masses in an SO(10) SUSY GUT with a single Higgs in the 10
representation \cite{Larry}. According this approach, only the third
family is allowed to receive mass from the renormalisable operators in
the superpotential.  The remaining masses and mixings are generated
from a minimal set of just three specially chosen non-renormalisable
operators whose coefficients are suppressed by some large scale.
Furthermore these operators are only allowed to contain adjoint 45
Higgs representations, chosen {}From a set of fields denoted $45_Y$,
$45_{B-L}$, $45_{T_{3R}}$, $45_X$ whose VEVs point in the direction of
the generators specified by the subscripts, in the notation of
\cite{Larry}.
This is precisely the strategy we wish to follow.  We shall
assume that only the third family receives its mass {}from a
renormalisable Yukawa coupling. All the other renormalisable Yukawa
couplings are set to zero. Then non-renormalisable operators are
written down which will play the role of small effective Yukawa
couplings. The effective Yukawa couplings are small because they
originate {}from non-renormalisable operators which are suppressed by
powers of the heavy scale $M$. We shall restrict
ourselves to all possible non-renormalisable operators which can be
constructed from different group theoretical contractions of the
fields:
\begin{equation} O_{ij}\sim (F_i\bar{F}_j
)h\left(\frac{H\bar{H}}{M^2}\right)^n+{\mbox h.c.} \label{op}
\end{equation} where we have used the fields $H,\bar{H}$ in
Eqs.\ref{H},\ref{barH} and $M$ is the large scale $M>M_{X}$.  The idea
is that when $H, \bar{H}$ develop their VEVs such operators will
become effective Yukawa couplings of the form $h F \bar{F}$ with a
small coefficient of order $M_X^2/M^2$. Although we
assume no intermediate symmetry breaking scale (i.e. SU(4)$\otimes
$SU(2)$_L \otimes $SU(2)$_R$ is broken directly to the standard model
at the scale $M_X$) we shall allow the possibility that there are
different higher scales $M$ which are relevant in determining the
operators.  For example one particular contraction of the indices of
the fields may be associated with one scale $M$, and a different
contraction may be associated with a different scale $M'$.  We shall
either appeal to this kind of idea in order to account for the various
hierarchies present in the Yukawa matrices, or to higher dimensional
operators which are suppressed by a further factor of $M$.

In the present model, although there are no adjoint representations,
there will in general be non-renormalisable operators which closely
resemble those in SO(10) involving adjoint fields.  The simplest such
operators correspond to $n=1$ in Eq.\ref{op}, with the $(H
\bar{H})$ group indices contracted together.

These operators are similar to those of ref.~\cite{Larry} but with
$H\bar{H}$ playing the r\^{o}le of the adjoint Higgs representations.
It is useful to define the following combinations of fields,
corresponding to the different $H\bar{H}$ transformation properties
under the gauge group in Eq.\ref{422},
\begin{eqnarray} (H \bar{H})_A & = & (1,1,1)
\nonumber \\ (H \bar{H})_B & = & (1,1,3)
\nonumber \\ (H \bar{H})_C & = & (15,1,1) \label{big} \\ (H \bar{H})_D
& = & (15,1,3)
\nonumber
\end{eqnarray}

It is straightforward to construct the
operators of the form of Eq.\ref{op} explicitly, and hence deduce the
effect of each operator.  For example for $n=1$ the four operators
are, respectively,
\begin{equation} O^{A,B,C,D}_{ij}\sim F_i\bar{F}_jh\frac{(H
\bar{H})_{A,B,C,D}}{M^2}+H.c.
\label{n1}
\end{equation} where we have suppressed gauge group indices.

These operators lead
to quark-lepton and isospin splittings, as shown
explicitly below:
\begin{eqnarray} O^{A}_{ij} & = & a_{ij} (Q_{i}\bar{U}_{j}h_2 +
Q_{i}\bar{D}_{j}h_1 + L_{i}\bar{E}_{j}h_1 + H.c.) \nonumber \\
O^{B}_{ij} & = & b_{ij} (Q_{i}\bar{U}_{j}h_2 - Q_{i}\bar{D}_{j}h_1 -
L_{i}\bar{E}_{j}h_1 + H.c.) \nonumber \\ O^{C}_{ij} & = & c_{ij}
(Q_{i}\bar{U}_{j}h_2 + Q_{i}\bar{D}_{j}h_1 -3L_{i}\bar{E}_{j}h_1 +
H.c.) \nonumber \\ O^{D}_{ij} & = & d_{ij} (Q_{i}\bar{U}_{j}h_2 -
Q_{i}\bar{D}_{j}h_1 + 3L_{i}\bar{E}_{j}h_1 + H.c.)
\label{effective}
\end{eqnarray} where the coefficients of the operators
$a_{ij},b_{ij},c_{ij},d_{ij}$ are all of order
$\frac{M_X^2}{M^2}$.

\section{Results and Conclusions}
Using operators such as those above, together with
more complicated $n=2$ operators, it is possible
to account for the entire fermion mass spectrum.
The successful ansatze \cite{422} involve 8
real parameters
plus an unremovable phase. With these 8
parameters we can describe the 13 physical
masses and mixing angles.
Third family Yukawa unification leads to a prediction for $m_t
(\mbox{pole}) = 130-200$ GeV and $\tan \beta = 35-65$, depending on
$\alpha_S (M_Z)$ and $\bar{m}_b$. More accurate predictions could be
obtained if the error on $\alpha_S(M_Z)$ and $\bar{m}_b$ were
reduced.
The analysis of the lower 2 by 2 block
of the Yukawa matrices leads to 2 possible
predictions for $\lambda_{\mu}/\lambda_s=3,4$ at the scale $M_X$
($3$ is the GJ prediction).
In the upper 2 by
2 block analysis we are led to 5 possible predictions for
$\lambda_{d}/\lambda_e=2,8/3,3,4,16/3$.
(again $3$ is the GJ prediction.) Finally, we have a
prediction that $|V_{ub}|>0.004$\cite{422}.

The high values of $\tan \beta$ required by our model (also predicted
in SO(10)) can be arranged by a suitable choice of soft SUSY breaking
parameters as discussed in ref.\cite{Carenaetal}, although this leads
to a moderate fine tuning problem \cite{Yuk}. The high value of $\tan
\beta$ is not stable under radiative corrections unless some other
mechanism such as extra approximate symmetries are invoked.  $m_t$ may
have been overestimated, since for high $\tan \beta$, the equations
for the running of the Yukawa couplings in the MSSM can get
corrections of a significant size from Higgsino--stop and
gluino--sbottom loops. The size of this effect depends upon the mass
spectrum and may be as much as 30 GeV. For our results to be
quantitatively correct, the sparticle corrections to $m_b$ must be
small. This could happen in a scenario with non-universal soft
parameters, for example. Not included in our analysis are threshold
effects, at low or high energies. These could alter our results by
several per cent and so it should be borne in mind that all of the
mass predictions have a significant uncertainty in them. It is also
unclear how reliable 3 loop perturbative QCD at 1 GeV is.

Despite a slight lack of predictivity of the model compared to SO(10), the
SU(4) $\otimes$SU(2)$_L \otimes$SU(2)$_R$ model has the twin
advantages of having no doublet-triplet splitting problem, and
containing no adjoint representations, making the model technically
simpler to embed into a realistic string theory. Although both these
problems can be addressed in the SO(10) model~\cite{new}, we find it
encouraging that such problems do not arise in the first place in the
SU(4)$\otimes$SU(2)$_L \otimes$SU(2)$_R$ model.  Of course there are
other models which also share these advantages such as flipped SU(5)
or even the standard model.  However, at the field theory level, such
models do not lead to Yukawa unification, or have precise Clebsch
relations between the operators describing the light fermion masses.
It is the combination of all of the attractive features mentioned
above which singles out the present model for serious consideration.


\newpage

\begin{thebibliography}{99}
\bibitem{Rabyi} S. Raby, Preprint No. OHSTPY-HEP-T-95-024; S. Raby,
Preprint No.  OHSTPY-HEP-T-95-001.
\bibitem{btauold} M. B. Einhorn and D. R. T. Jones, Nucl. Phys. {\bf
B196} (1982) 475; J. Ellis, D. V. Nanopoulos and S. Rudaz,
Nucl. Phys. {\bf B202} (1982) 43.
\bibitem{btaunew} B.~C. Allanach and S.~F. King,
\newblock Phys. Lett. {\bf B328}, (1994) 360; H. Arason {\it et al},
\/Phys. Rev. Lett. {\bf 67} (1991) 2933; {\it ibid} \/Phys. Rev. {\bf
D47} (1993) 232; P. Langacker and N. Polonsky, UPR-0556T.  V.Barger,
M.S.Berger and P.Ohmann,
\newblock Phys. Rev. {\bf D47}, 1093 (1993).
\bibitem{GJ} H. Georgi and C. Jarlskog, Phys. Lett. {\bf B86} (1979)
297.
\bibitem{DHR} S. Dimopoulos, L. Hall and S. Raby, Phys. Rev. {\bf D45}
(1992) 4192.
\bibitem{frogetal} C. D. Froggatt and H. B. Nielson, Nucl. Phys. {\bf
B147} (1979) 277; Z. G. Berezhiani, Phys. Lett. {\bf B129} (1983) 99;
ibid., {\bf B150} (1985) 177; S. Dimopoulos, Phys. Lett. {\bf B129}
(1983) 417.
\bibitem{RRR} P. Ramond, R. Roberts and G. Ross, Nucl. Phys. {\bf
B406} (1993) 19.
\bibitem{pati} J. Pati and A. Salam, Phys. Rev. {\bf D10} (1974) 275.
\bibitem{leo1} I. Antoniadis and G. K. Leontaris, Phys. Lett. {\bf
B216} (1989) 333.
\bibitem{leo2} I. Antoniadis, G. K. Leontaris and J. Rizos,
Phys. Lett. {\bf B245} (1990) 161.
\bibitem{flipped} I. Antoniadis, J. Ellis, J. Hagelin and
D. Nanopoulos, Phys. Lett. {\bf B194} (1987) 231.
\bibitem{Larry} G. Anderson, S. Raby, S. Dimopoulos, L. Hall and
G. Starkman, Phys. Rev {\bf D49} (1994) 3660
\bibitem{422} S. F. King Phys. Lett. {\bf B325} (1994) 129; \\
B. C. Allanach and S. F. King, hep-ph/9502219,
submitted to Nucl. Phys. {\bf B.}
\bibitem{Yuk} B. Ananthanarayan, G. Lazarides and Q. Shafi,
Phys. Rev. {\bf D44} (1991) 1613; H. Arason {\it et al},
Phys. Rev. Lett. {\bf 67} (1991) 2933; L. Hall, R. Rattazzi and
U. Sarid, LBL preprint 33997.
\bibitem{Carenaetal}
M. Carena, M. Olechowski, S. Pokorski and C. E. M. Wagner,
Nucl. Phys. {\bf B426} (1994) 269.
\bibitem{new} S. Chaudhuri, S. Chung and J. Lykken, FERMILAB
PUB-94/137-T; L. Hall and S. Raby, OHSTPY-HEP-T-94-023, LBL-36357,
UCB-PTH-94/27.
\newblock

\end{thebibliography}
\end{document}